\documentclass[%
reprint,
superscriptaddress,
showpacs,
nofootinbib,
amsmath,
amssymb,
amsfonts,
aps,
pra,
floatfix,
showkeys
]{revtex4-2}

\RequirePackage{atbegshi}

\usepackage[utf8]{inputenc}
\usepackage[T1]{fontenc}
\usepackage[ngerman,english]{babel}
\usepackage[matha]{mathabx} 
\usepackage[dvipsnames, svgnames, table]{xcolor}
\usepackage{%
    graphicx,
    siunitx,
    multirow,
    booktabs,
    bbold,    
    pifont,   
    capt-of,
    pgfplotstable,
    todonotes,
}
\usepackage{subfigure}     
\renewcommand{\thesubfigure}{(\alph{subfigure})}
\makeatletter
  \renewcommand{\@thesubfigure}{\thesubfigure\space}
  \def\@currentlabel{\p@subfigure\thesubfigure}
\makeatother
\usepackage{hyperref}
\usepackage{braket}
\hypersetup{
    colorlinks=true,
    citecolor=black,
    linkcolor=black,
    urlcolor=black,
    anchorcolor=black,
    linktocpage
}
\usepackage{outlines}
\usepackage{etoolbox} 
\usepackage{cleveref} 
\makeatletter
\appto{\appendix}{%
  \@ifstar{\def\theequation@prefix{A.}}%
          {}%
}
\makeatother
\crefname{figure}{Figure}{Figures}
\crefname{table}{Table}{Tables}
\crefname{equation}{Eq.}{Eqs.}
\crefname{section}{Section}{Sections}
\crefformat{appendix}{the #2Appendix#1#3} 
\graphicspath{{./figures/}}

\DeclareMathOperator{\Tr}{Tr}

\newcommand{\clqcd}{CL\kern-.25em\textsuperscript{2}QCD}
\newcommand{\HE}{\hat{H}_{\text{E}}}
\newcommand{\HB}{\hat{H}_{\text{B}}}
\newcommand{\HKS}{\hat{H}_{\text{KS}}}
\newcommand{\tj}[6]{ \begin{pmatrix}
  #1 & #2 & #3 \\
  #4 & #5 & #6 
\end{pmatrix}}

\newcommand{\beq} {\begin{eqnarray}}
\newcommand{\eeq} {\end{eqnarray}}
\newcommand{\nn}{ \nonumber}

\begin{document}

\title{
Dihedral Lattice Gauge Theories on a Quantum Annealer 
}

\author{Michael Fromm}
 \email{mfromm@itp.uni-frankfurt.de}
 \affiliation{
  Institut f\"{u}r Theoretische Physik, Goethe-Universit\"{a}t Frankfurt\\
 Max-von-Laue-Str.\ 1, 60438 Frankfurt am Main, Germany
}

\author{Owe Philipsen}
 \email{philipsen@itp.uni-frankfurt.de}
 \affiliation{
  Institut f\"{u}r Theoretische Physik, Goethe-Universit\"{a}t Frankfurt\\
 Max-von-Laue-Str.\ 1, 60438 Frankfurt am Main, Germany
}

\author{Christopher Winterowd}
\email{winterowd@itp.uni-frankfurt.de}
\affiliation{
 Institut f\"{u}r Theoretische Physik, Goethe-Universit\"{a}t Frankfurt\\
 Max-von-Laue-Str.\ 1, 60438 Frankfurt am Main, Germany
}

\begin{abstract}
We study lattice gauge theory with discrete, non-Abelian gauge groups. We extend the formalism of previous studies on D-Wave's quantum annealer as a computing platform to finite, simply reducible gauge groups. As an example, we use the dihedral group $D_n$ with $n=3,4$ on a two plaquette ladder for which we provide proof-of-principle calculations of the ground-state and employ the known time evolution formalism with Feynman clock states.
\end{abstract}

\maketitle

\section{Introduction}

Lattice gauge theory (LGT) calculations using quantum computers have already seen substantial progress. This is despite the fact that programmable quantum hardware has only recently become widely available to researchers in physics (see e.g.~\cite{Bauer2022} for an up-to-date high-energy physics perspective on the field). With the basic formulation of quantum simulations of LGT being laid out very early on~\cite{ByrnesYamamoto2005}, efficient formulations of abelian~\cite{Haase2020} and non-abelian LGT~\cite{Klco2020, Klco2021} on universal, gate-based hardware now exist. This includes a complete set of instructions for the efficient and accurate simulation of QCD and QED~\cite{Kan2021}. 

On the side of adiabatic quantum computing~\cite{RevModPhys.90.015002}, despite the fact that it has been commercially available for more than a decade in the form of quantum annealers (QA)~\cite{Johnson_2011}, this approach has only recently been used for LGT calculations. These include the pioneering studies on the annealer for the case of $SU(2)$~\cite{Lewis2021} and $SU(3)$~\cite{Illa2022}. In these formulations, the number of qubits necessary to digitize the theory under study scales with the size of the Hilbert space of the problem rather than the spatial volume of the system. Thus, this formulation does not show the expected quantum advantage present in universal quantum computing. On the other hand, systems in a QA architecture s.a. D-Wave's {\tt Advantage\_system5.1} already comprise several thousand physical qubits~\cite{Dwave}. Even at this stage of hardware development, proof-of-principle quantum computations in LGT~\cite{Lewis2021, Illa2022} or other field theories~\cite{Abel2020} are feasible. The intrinsic nature of the formulation of problems on the annealer simply requires the mapping of the lattice field theory onto an optimization problem represented by the Hamiltonian
\beq \label{eq:qubo_form}
H(q) = \sum_i Q_{ii}q_i + \sum_{i<j} Q_{ij}q_i q_j\,,
\eeq 
with a real, upper-triangular matrix $Q$ and binary variables $q_i\in\{0,1\}$. This type of problem goes under the name of quadratic unconstrained binary optimization (QUBO). Thus, QA can be seen as an ideal entry point into the field of quantum simulation of LGT. It is in this spirit, that we turn to the subject of our study.

On the lattice, the gauge field Hilbert space is infinite for compact Lie groups and a truncation of the full symmetry becomes necessary (see~\cite{ZoharChallenges2021} for a basic overview). One such truncation is the approximation of the full symmetry group by a discrete subgroup, a strategy whose varied success and intricacies are well summarized in~\cite{Alam2021}. Here, we take the practical approach by choosing the finite, non-abelian dihedral group $D_n$ as our gauge group, which can be digitized without truncation as in~\cite{Lamm2019}. By mapping this problem onto an optimatization problem amenable to the QA, we extend the previous studies of compact Lie groups~\cite{Lewis2021, Illa2022} to the case of simply reducible, finite groups for which we provide the adapted framework.
\section{Hamiltonian $D_n$ lattice gauge theory}
We begin by introducing the Hamiltonian formulation of $D_n$ lattice gauge theory. Our approach closely follows that of ~\cite{ByrnesYamamoto2005}, where the Hamiltonian approach was worked out for a general gauge group $G$. As usual, we work on a cubic lattice of dimension $d$, with the gauge fields living on the links between the lattice sites. The gauge field Hilbert space, denoted by $\mathcal{H_G}$, is a direct product of the individual link spaces $\mathcal{H_G} = \bigotimes_{\ell}\mathcal{H}_\ell $ which, in the group element basis, are defined by $\mathcal{H}_\ell = \mathrm{span}(\{\ket{g}\}_{g\in G})$. We start by defining the link operator $\hat{U}$, acting on $\mathcal{H}_\ell$, as
\beq \label{eq:U_def}
\hat{U}^j_{mn} = \int\mathrm{dg}\,D^j_{mn}(g) \ket{g}\bra{g}
\eeq
where $j$ labels the irreducible representations (irreps) of $G$ and $D^j_{mn}(g)$ are the Wigner representation matrices for $g\in G$. The indices $m,n$ label the multiplicity of the left and right projection of the link, respectively. This object is of primary importance in the Hamiltonian formulation as it is responsible for the interactions.

\begin{figure}
\includegraphics[scale=.25]{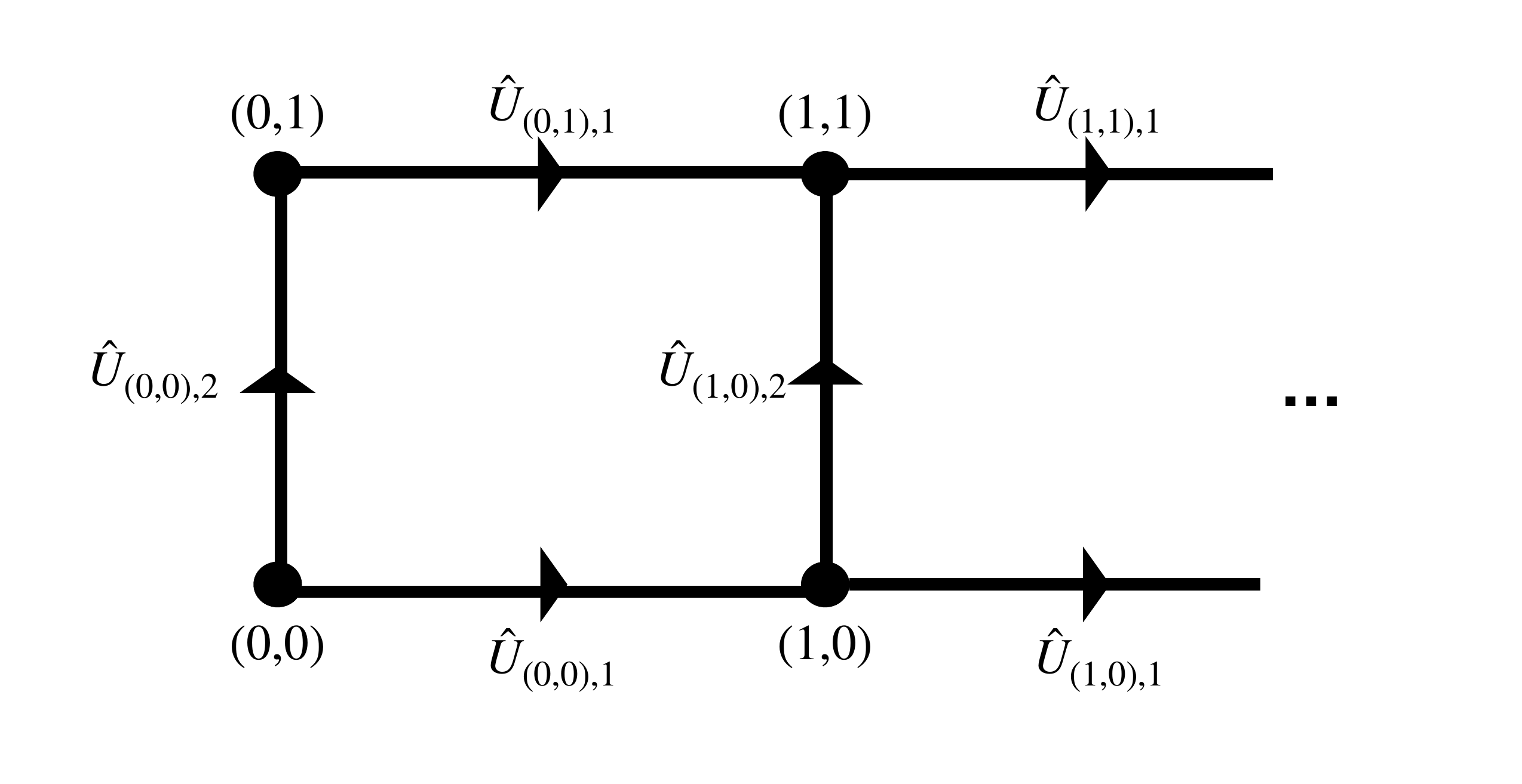}
\caption{\label{fig:ladder} Ladder geometry used in this study. The sites of the ladder have been labeled as $(i,j)$, where $i=0,1,\dots,N-1$ and $j=0,1$. Here $N$ denotes the length of the ladder. The forward link operators are also shown and have been labeled by their direction $\mu=1,2$ as well as the site from which they emanate. We note that the system is periodic only in the $\hat{1}$-direction.}
\end{figure}

The Hamiltonian formulation for $G=SU(2)$ lattice gauge theory was first written down in \cite{KogutSusskind}. In later work, it was shown how this Hamiltonian could be obtained from the transfer matrix \cite{PhysRevD.15.1128}. For a general gauge group $G$, the lattice Hamiltonian, commonly referred to as the Kogut-Susskind Hamiltonian, consists of two terms
\beq \label{eq:HKS_sum}
\HKS = \HE + \HB, 
\eeq 
where the first term is referred to as the electric part and the second term is referred to as the magnetic part. The magnetic part takes the form 
\beq \label{eq:KS_Hamiltonian_magnetic}
\HB = -\lambda_B\sum_{\vec{x},i<j}\mathrm{Re}\Tr \hat{U}_i(\vec{x}) \hat{U}_j(\vec{x}+\hat{i}) \hat{U}^{\dagger}_i(\vec{x}+\hat{j}) \hat{U}^{\dagger}_j(\vec{x}),\nn\\
\eeq 
 where the sum is over the spatially-oriented plaquettes\footnote{For the case of $G=D_3,D_4$ which we will consider in this work, we take the link operators to be in the faithful, two-dimensional irrep, which we denote by $\mathbb{2}$.} and can be seen to be diagonal in the group elements basis by virtue of the definition in Eq.(\ref{eq:U_def}). In this basis, the electric term, $\HE$, has a form that depends on whether $G$ is taken to be a compact Lie group or a finite group \cite{HarlowOoguri2018, PhysRevD.104.094519}. It is often much more convenient to work in the representation basis, labeled by the states $\ket{jmn}$. Here $j$ labels the irrep and $m,n$ run over the states within the multiplet. In this basis $\HE$ becomes diagonal
\beq \label{eq:KS_Hamiltonian_electric}
\HE = \lambda_E \sum_{\vec{x}} \sum^d_{i=1} \sum_{jmn} f_j \ket{jmn}_{\vec{x},i}\bra{jmn}_{\vec{x},i}\,.
\eeq 
The group element basis is related to the representation basis through the following relation
\beq \label{eq:representation_basis_group_basis_overlap}
\braket{g|jmn} = \sqrt{\frac{\text{dim}(j)}{|G|}} D^j_{mn}(g),
\eeq
where $|G|$ is the order of the finite group $G$. From here on out, we will assume that we are working with finite groups and our formulae will reflect this.
Using the transformation in Eq.(\ref{eq:representation_basis_group_basis_overlap}), one can easily transform between the two bases. 

It is important to discuss the couplings in the Kogut-Susskind Hamiltonian.
The relationship between the couplings $\lambda_E, \lambda_B$ appearing in front of the electric and magnetic terms, respectively, can be determined in the limit $a_0\to 0$ when deriving the Hamiltonian in the transfer matrix formalism. However, this procedure depends on the gauge group under consideration. While for a compact Lie group, $\lambda_E = g_H^2/2,\lambda_B = 1/g_H^2$, for a finite group it has been determined by previous studies that one should use $\lambda_E = \exp{(-2/g^2_H)},\lambda_B = 1/g_H^2$, instead \cite{HarlowOoguri2018, PhysRevD.104.094519}. Here we have introduced the Hamiltonian coupling $g_H$, which is the geometric mean of the spatial and temporal couplings, $g_s$ and $g_t$. These are introduced in the transfer matrix formulation when one takes the temporal and spatial lattice spacings to be distinct. The coefficients $f_j$ appearing in Eq.(\ref{eq:KS_Hamiltonian_electric}) are the eigenvalues of the quadratic Casimir operator for the case of compact Lie groups \cite{ByrnesYamamoto2005}. For example, when $G = SU(2)$, one gets the familiar result $f_j = j(j+1)$. On the other hand, for finite gauge groups, the $f_j$ can be derived systematically from the transfer matrix in the limit of vanishing temporal lattice spacing, $a_0\to 0$~\cite{PhysRevD.104.094519}. Other choices for the coefficients for the case of discrete gauge groups also exist in the literature~\cite{Bender2018}.
\subsection{Computation of $H_{ij}$}
With the Kogut-Susskind Hamiltonian given in operator form Eq.(\ref{eq:HKS_sum}), there still remains the task of constructing the states of the physical Hilbert space $\mathcal{H_P}$. The label physical here refers to the subspace of the larger Hilbert space $\mathcal{H_G}$ that respects local gauge invariance. Formally, in the group element basis, local gauge invariance can be expressed with the help of left and right multiplication operators given by $\Theta^L_g(\vec{x}, i)$ and $\Theta^R_g(\vec{x}, i)$ that act as follows
\beq 
\Theta^L_g \ket{h} = \ket{g^{-1}h}, ~
\Theta^R_g \ket{h} = \ket{hg^{-1}}, ~g,h \in G,
\eeq 
with $\Theta^{R \dagger}_g = \Theta^R_{g^{-1}}$. As usual, the link transforms in the adjoint representation under a gauge transformation. Thus, a local gauge transformation at the site $\vec{x}$ parametrized by $g$ is given by
\beq 
\Tilde{\Theta}_g(\vec{x}) \equiv \prod^{d}_{i=1} \Theta^L_g(\vec{x},i) \Theta^{R \dagger}_g(\vec{x}-\hat{i},i). 
\eeq 
For a generic physical state $\ket{\psi}$, gauge invariance demands
\beq
\label{eq:local_gauge_invariance}
\Tilde{\Theta}_g(\vec{x}) \ket{\psi} = \ket{\psi}, \forall~\vec{x}\in \mathbb{Z}^d.
\eeq 
In the representation basis, the statement of Gauss's law in Eq.(\ref{eq:local_gauge_invariance}) is equivalent to $\ket{\psi}$ being written as a direct product of color singlets at each lattice site
\beq
\ket{\psi} = \bigotimes_{\vec{x}} \ket{00}_{\vec{x}}\,,
\eeq
where we refer to Appendix~\ref{section:Gauss} for the details rearding the explicit construction of $\ket{00}_{\vec{x}}$ in terms of the $\ket{jmn}$.

As we are ultimately interested in mapping our system onto a quantum annealer, we are restricted to a rather small system size (Fig.~(\ref{fig:ladder})). For a given lattice geometry, we are then left with the task of determining the matrix elements of the Hamiltonian (\ref{eq:HKS_sum}). We start by introducing the trivial vacuum state $\ket{0}$ with every link $\ell$ in the trivial representation, $j_\ell = 0$, s.t. $\HE \ket{0} = 0$. Physically, this corresponds to the situation where the links contain zero chromoelectric flux. Following the approach of \cite{ByrnesYamamoto2005}, the states of the physical Hilbert space can be systematically generated from this configuration by subsequently acting with gauge-invariant operators. To carry out this procedure, one needs to know how an individual link operator acts on a general state in the representation basis, $\ket{jmn}$. Using the definition of the link operator in (\ref{eq:U_def}) and the matrix element in (\ref{eq:representation_basis_group_basis_overlap}), one obtains
\beq \label{eq:2irrep_on_rep_basis_ket_pt1}
\hat{U}^{\mathbb{2}}_{m'n'} \ket{jmn} = \sqrt{\frac{\text{dim}(j)}{|G|}} \sum_{g \in G} D^2_{m'n'}(g) D^j_{mn}(g)\ket{g}\,.\nn\\
\eeq 
This can be further simplified by using the Clebsch-Gordan (CG) series for the tensor product of two arbitrary representation matrices, which yields the result
\beq
 \label{eq:2irrep_on_rep_basis_ket_pt2}
\hat{U}^{\mathbb{2}}_{m'n'} \ket{jmn} &=& \sum_{J} \sum_{M} \sum_{N} \sqrt{\frac{\text{dim}(j)}{\text{dim}(J)}}  \\ 
 &\times& \braket{2m'jm|JM} \braket{JM|2n'jn} \ket{JMN},\nn
\eeq 
where the sum on $J$ is over the irreps and the sums over $M$ and $N$ are over the states in a given irrep.\footnote{We note that for our choice of the $\mathbb{2}$ irrep of $D_3$ and $D_4$ Eq.(\ref{eq:two_dim_rep_standard}), the link operators satisfy $U^{\mathbb{2}\dagger}_{m,n} = U^{\mathbb{2}}_{n,m}$.}

Applying gauge invariant operators repeatedly to the trivial vacuum state with the help of Eq.(\ref{eq:2irrep_on_rep_basis_ket_pt2}), the enumeration of the configuration space can be performed. In this procedure, Gauss's law is imposed at each step. This is equivalent to the Wigner $3$J-symbol being nonzero for a tuple of irreps, $(j_1,j_2,j_3)$, which characterize the three links involving a given lattice site. For more details regarding the $3$J symbols and Gauss's law we refer the reader to appendices (\ref{subsection:3J}) and (\ref{section:Gauss}).

This task of mapping out the physical Hilbert space can be automated using a Markov-chain-like approach. For this, all one needs are the $3$J symbols for the given gauge group $G$. The total number of states in the full Hilbert space is $\tilde{N}_{\text{irreps}}^{3N}$ states, where $\tilde{N}_{\text{irreps}}$ is the total number of irreps of the gauge group $G$ and $N$ is the size of the ladder. Although enforcing local gauge invariance removes a large number of these states, the physical Hilbert space still grows quite rapidly. In Table \ref{table:counts_configs}, we display the size of the physical Hilbert space, where $G = D_3, D_4$. These counts show that even for modest system sizes and small non-abelian groups there are constraints as to the problems that can be mapped to the quantum annealer. 
\begin{table}[]
\centering
\begin{tabular}{|c| c c c |} 
\hline
  $N$ & 2 & 3 & 4
 \\ [0.5ex] 
 \hline
 $D_3$ & 49 & 251 & $O(1300)$ \\ 
  $D_4$ & 76 &  392 &  $O(2500)$ \\  [1ex] 
 \hline
\end{tabular}
\caption{List of the size of the physical Hilbert space, $N_{\text{conf}}$, on a ladder of size $N$ for $D_3$ and $D_4$. The configurations are enumerated by a set of integers $\{j_i, i=1,2,\dots,3N \}$ characterizing the irrep of each link on whereby Gauss's law is satisfied at each site.}
\label{table:counts_configs}
\end{table}

Once we have enumerated the states in the physical Hilbert, space we can finally compute the matrix elements of the Hamiltonian in this basis. As $\HE$ is diagonal in the representation basis, the application of Eq.(\ref{eq:KS_Hamiltonian_electric}) to the generic state in (\ref{eq:generic_gauss_psi}) is trivial. The magnetic Hamiltonian, which is responsible for the interactions, has a nontrivial action on a physical state. 
To illustrate this, we first write an arbitrary plaquette operator on the ladder
\beq \nn
\hat{P}_{\vec{x}} &=& \sum_{n_1,\ldots,n_4} \hat{U}_{\vec{x},1;n_1,n_2} \hat{U}_{\vec{x}+\hat{1},2;n_2,n_3} \\  
&\times& \hat{U}_{\vec{x}+\hat{2},1;n_3,n_4}^{\dagger}\hat{U}_{\vec{x},1;n_4,n_1}^{\dagger},
\label{eq:arb_plaquette_ladder}
\eeq 
where $\vec{x} = (x,0), ~ x=0,\ldots,N-1$ denotes the vertex at the bottom left corner of the plaquette and the sum is over the group indices. Here we label each link by its site vector and direction. Now, using the relation Eq.(\ref{eq:2irrep_on_rep_basis_ket_pt2}), we can determine the result of the plaquette operator acting on a state. The matrix element of Eq.(\ref{eq:arb_plaquette_ladder}) between two arbitrary physical states is given by
\beq \nn 
\bra{\psi'}\hat{P}_{\vec{x}}\ket{\psi} &=& \sum_{n_1,\ldots,n_4}\sum_{\{m_{i,1},o_{i,1}\}}\cdots\sum_{\{m_{i,n_s}, o_{i,n_s}\}} \\ \nn &\times& \prod_s\tj{j_{\ell_{1,s}}}{j_{\ell_{2,s}}}{j_{\ell_{3,s}}}{m_{1,s}}{m_{2,s}}{m_{3,s}}  \overline{\tj{l_{\ell_{1,s}}}{l_{\ell_{2,s}}}{l_{\ell_{3,s}}}{o_{1,s}}{o_{2,s}}{o_{3,s}}} \\ \nn &\times& \prod_{i \not\in \mathcal{L}_{\vec{x}}} \delta_{l_i,j_i} \delta_{m_{R_i},o_{R_i}} \delta_{m_{L_i},o_{L_i}} \\ \label{eq:arb_plaquette_matrix_element}
&\times& \frac{\sqrt{\mathrm{dim}(j_{\tilde{l}_1}) \mathrm{dim}(j_{\tilde{l}_2}) \mathrm{dim}(j_{\tilde{l}_3}) \mathrm{dim}(j_{\tilde{l}_4})}}{\sqrt{\mathrm{dim}(l_{\tilde{l}_1}) \mathrm{dim}(l_{\tilde{l}_2}) \mathrm{dim}(l_{\tilde{l}_3}) \mathrm{dim}(l_{\tilde{l}_4})}} \\ \nn
&\times& \braket{2 n_1 j_{\tilde{l}_1} m_{L_{\tilde{l}_1}} | l_{\tilde{l}_1} o_{L_{\tilde{l}_1}}}\braket{l_{\tilde{l}_1} o_{R_{\tilde{l}_1}} | 2 n_2 j_{\tilde{l}_1} m_{R_{\tilde{l}_1}} } \\ \nn 
&\times& \braket{2 n_2 j_{\tilde{l}_2} m_{L_{\tilde{l}_2}} | l_{\tilde{l}_2}  o_{L_{\tilde{l}_2}}}\braket{l_{\tilde{l}_2} o_{R_{\tilde{l}_2}} | 2 n_3 j_{\tilde{l}_2} m_{R_{\tilde{l}_2}} } \\ \nn 
&\times& \braket{2 n_4 j_{\tilde{l}_3} m_{L_{\tilde{l}_3}} | l_{\tilde{l}_3} o_{L_{\tilde{l}_3}}}\braket{l_{\tilde{l}_3} o_{R_{\tilde{l}_3}} | 2 n_3 j_{\tilde{l}_3} m_{R_{\tilde{l}_3}} } \\  \nn
&\times& \braket{2 n_1 j_{\tilde{l}_4} m_{L_{\tilde{l}_4}} | l_{\tilde{l}_4} o_{L_{\tilde{l}_4}}}\braket{l_{\tilde{l}_4} o_{R_{\tilde{l}_4}} | 2 n_4 j_{\tilde{l}_4} m_{R_{\tilde{l}_4}} },
\eeq 
where we refer to $\ket{\psi}$ and $\ket{\psi'}$ as the ``in" and ``out" states, $\mathcal{L}_{\vec{x}} = \{ \tilde{l}_1,  \tilde{l}_2,  \tilde{l}_3,  \tilde{l}_4 \}$ denotes the set of all links involved in (\ref{eq:arb_plaquette_ladder}), and the bar denotes complex conjugation. The sums over $n_i$ run over the states in the $\mathbb{2}$ representation and the sums over $m_{i,\mu}, o_{i,\mu}$ run over the states in the corresponding irreps for each link in the ``in" and ``out" states.  In Eq.(\ref{eq:arb_plaquette_matrix_element}), we have introduced the Wigner $3$J symbols which can be generalized to $D_n$. This along with other details regarding the group theory of $D_n$ are given in appendix (\ref{sec:appendix_A}), with the full derivation of Eq.(\ref{eq:arb_plaquette_matrix_element}) in appendix (\ref{sec:plaq_contrib}). Using this result for the plaquette matrix element, one can construct the magnetic Hamiltonian by recalling that $\HB = -\lambda_B \sum_{\vec{x}} \hat{P}_{\vec{x}}$. We note here that this construction is completely general and applies to a ladder of arbitrary length. 
\begin{figure}
\includegraphics[width=.47\linewidth]{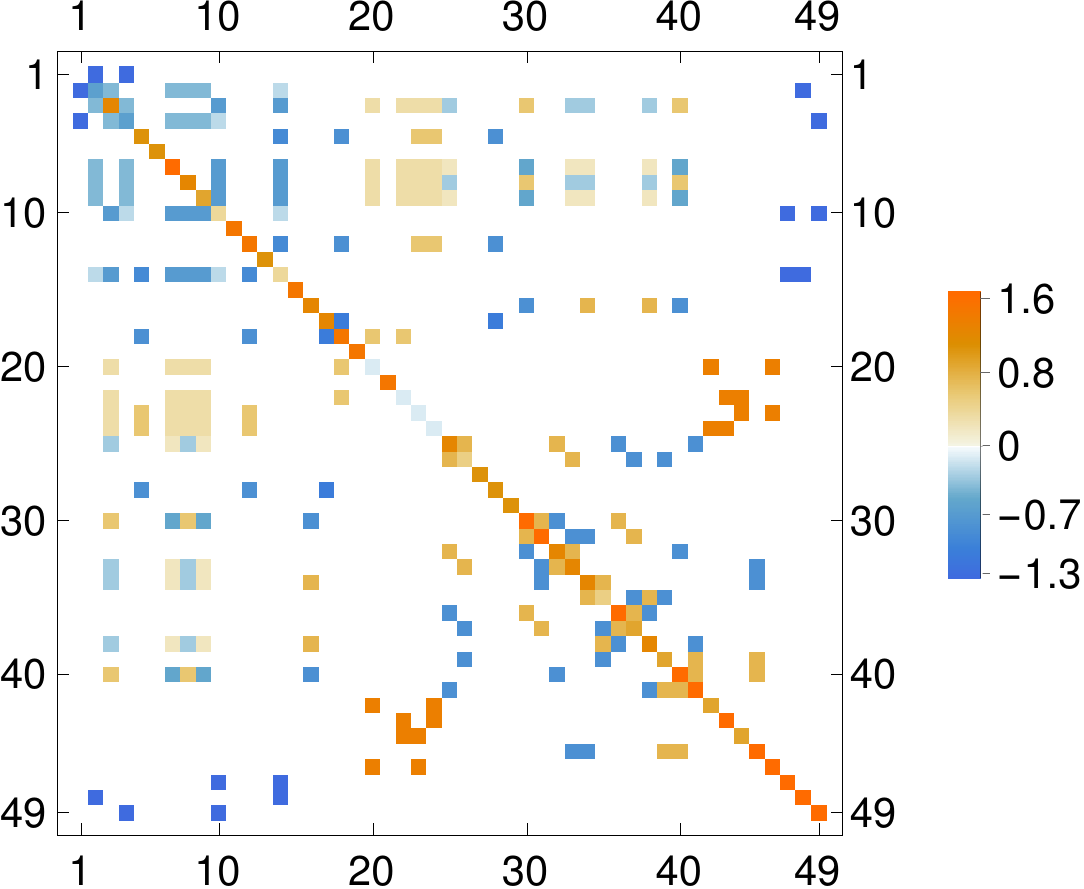}
\includegraphics[width=.47\linewidth]{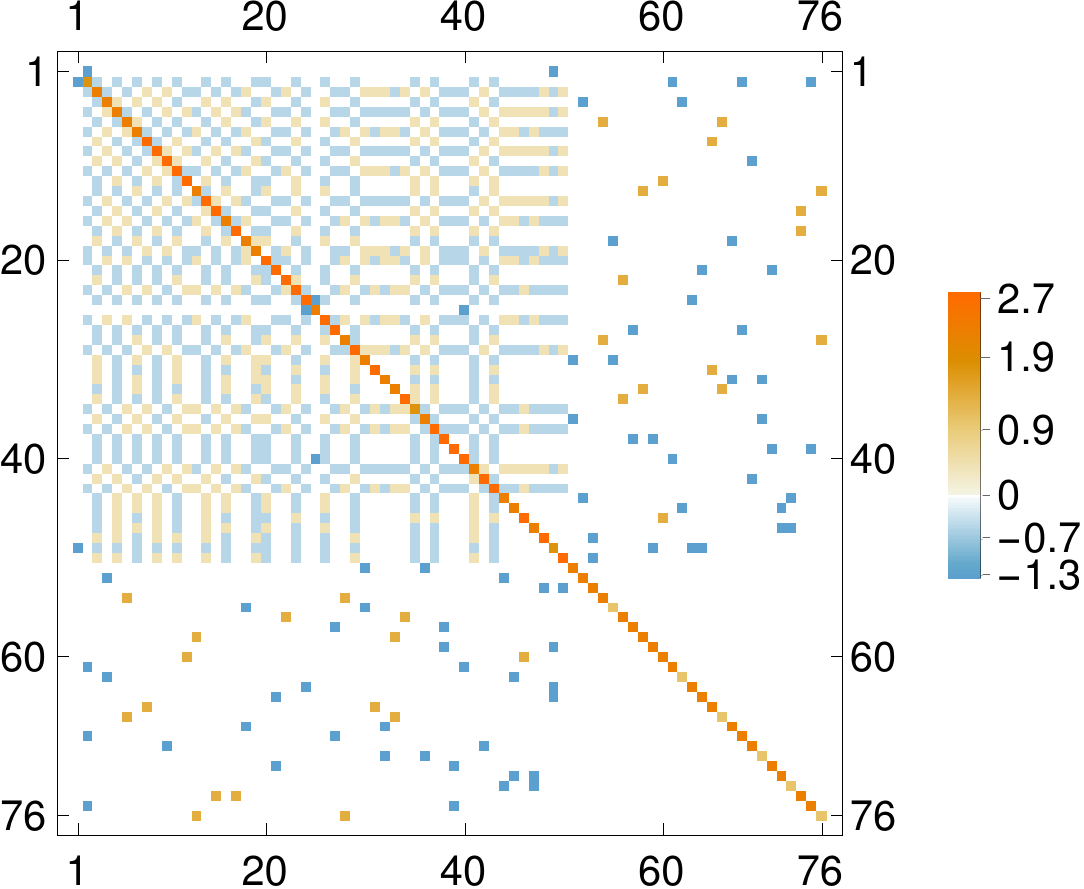}
\caption{\label{fig:dn_hamiltonian} Visualization of the Hamiltonian matrices ($g_H^2 = 0.75$) for gauge groups $D_3$ (left) and $D_4$ (right) on the $N=2$ ladder. One immediately notices the sparsity of both Hamiltonians, which is due to the structure of the magnetic contribution.}
\end{figure}

For the gauge groups which we have examined, it turns out that the Hamiltonian matrix is extremely sparse. This will work to our advantage later on when we map our problem to the quantum annealer. In Fig.\ref{fig:dn_hamiltonian}, we display a visualization of the sparsity of the Hamiltonian for both $D_3$ and $D_4$. Examining the product of delta functions in Eq.(\ref{eq:arb_plaquette_matrix_element}), the sparseness of the Hamiltonian ultimately comes from the orthonormality of the link states in the representation basis.
Once one has calculated the Hamiltonian matrix, the classical part of the calculation is practically complete. In the following, we discuss how our lattice gauge Hamiltonian is transformed into an optimization problem so that the quantum computation on the annealer can be performed.
\section{Implementation and Results}
\subsection{Groundstate via variational formulation}
Quantum annealing is a method used to solve a very specific type of problem: the calculation of the ground state of a generalized Ising model~\cite{Dwave}. Thus, unlike the case of the gate-based approach where one has at one's disposal a set of universal quantum gates, for quantum annealing one must cast the problem that one would like to solve into the form of an Ising model.

In the context of lattice gauge theory, it has been shown that one can map the Hamiltonian in Eq.(\ref{eq:HKS_sum}) onto a model with QUBO form, Eq.(\ref{eq:qubo_form}), in order to compute the low-lying states of the spectrum \cite{Illa2022, Lewis2021}. 
To see how this emerges we consider the variational principle from quantum mechanics
\beq
\label{eq:variational_bound}
E_0 \leq \frac{\bra{\psi} \hat{H}\ket{\psi}}{\braket{\psi|\psi}}\,,
\eeq
where $E_0$ is the ground-state energy.
Here we use a variational ansatz with a trial wave function 
\beq \label{eq:variational_ansatz}
\ket{\psi} = \sum_{\alpha=1}^{N_{\mathrm{conf}}} a_{\alpha} \ket{\phi_\alpha},
\eeq
with real parameters $a_\alpha$ and basis states $\ket{\phi_\alpha}$ corresponding to the configurations of the physical Hilbert space. The expansion parameters are, in general, complex but here can be chosen to be real as the Hamiltonian is a real, symmetric matrix. It is in this way that solving for the ground-state energy of our lattice Hamiltonian can be recast as an optimization problem as one seeks to minimize $\bra{\psi} \hat{H} \ket{\psi}$. Our accuracy in determining the eigenstates of the system is limited only by the precision to which we can determine the coefficients $a_{\alpha}$. To cast our problem in the form of Eq.(\ref{eq:qubo_form}), we do the following. First, the coefficients are given a fixed-point binary representation. Second, the norm of the wave function is discouraged from being zero by adding a penalty term. Incorporating both of these into our variational calculation, the \text{rhs} of Eq.(\ref{eq:variational_bound}) can rewritten as a cost function 
\beq
\label{eq:cost_function}
F = \bra{\psi} \hat{H} \ket{\psi} - \eta \braket{\psi|\psi} = \sum_{\alpha,\beta}^{{N_\mathrm{conf}}}\sum_{i,j}^K Q_{\alpha\beta,ij} q_{\alpha,i}q_{\beta,j},\nonumber\\
\eeq

\begin{figure*}[ht]
\includegraphics[scale=0.4]{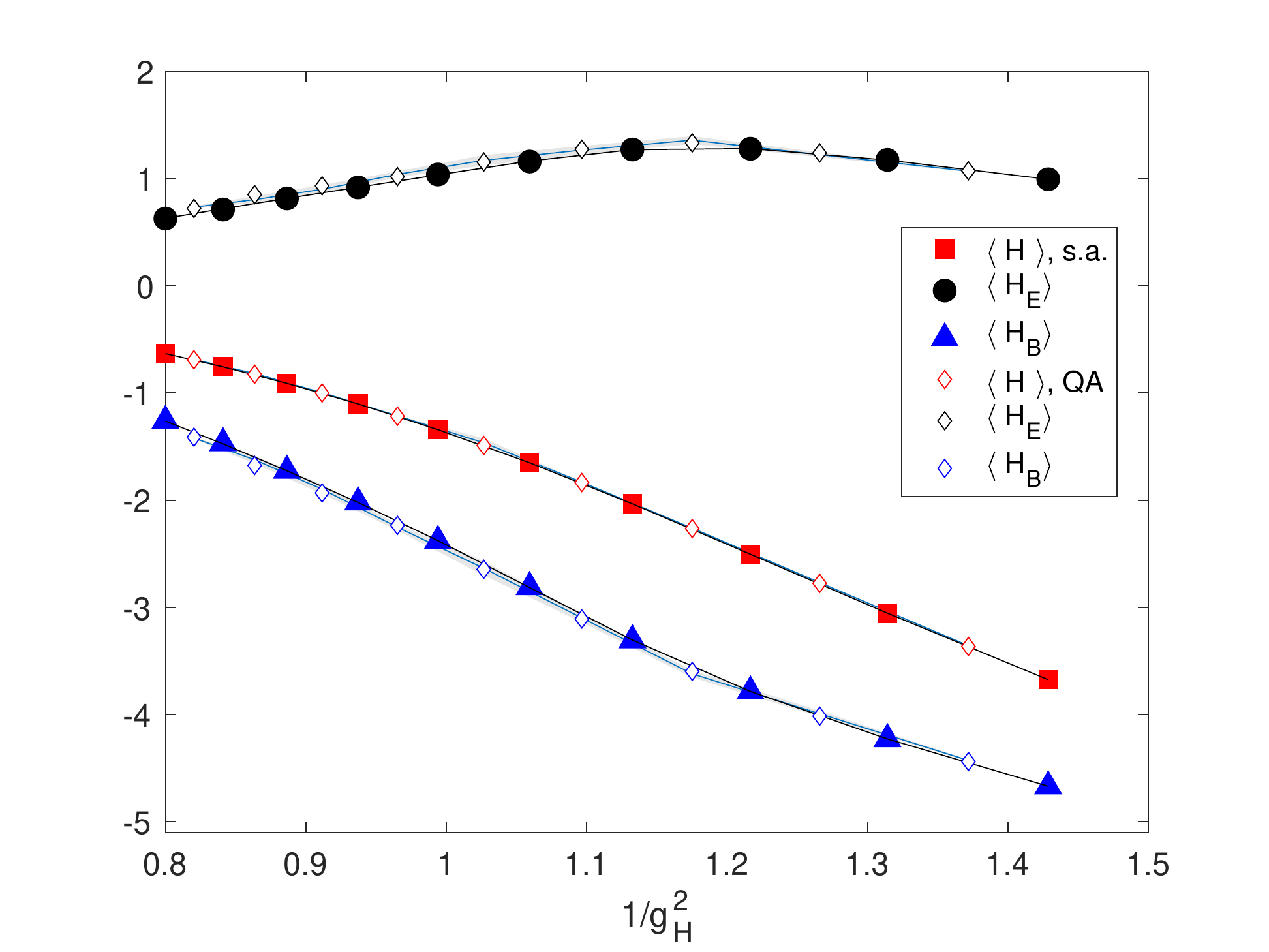}
\includegraphics[scale=0.4]{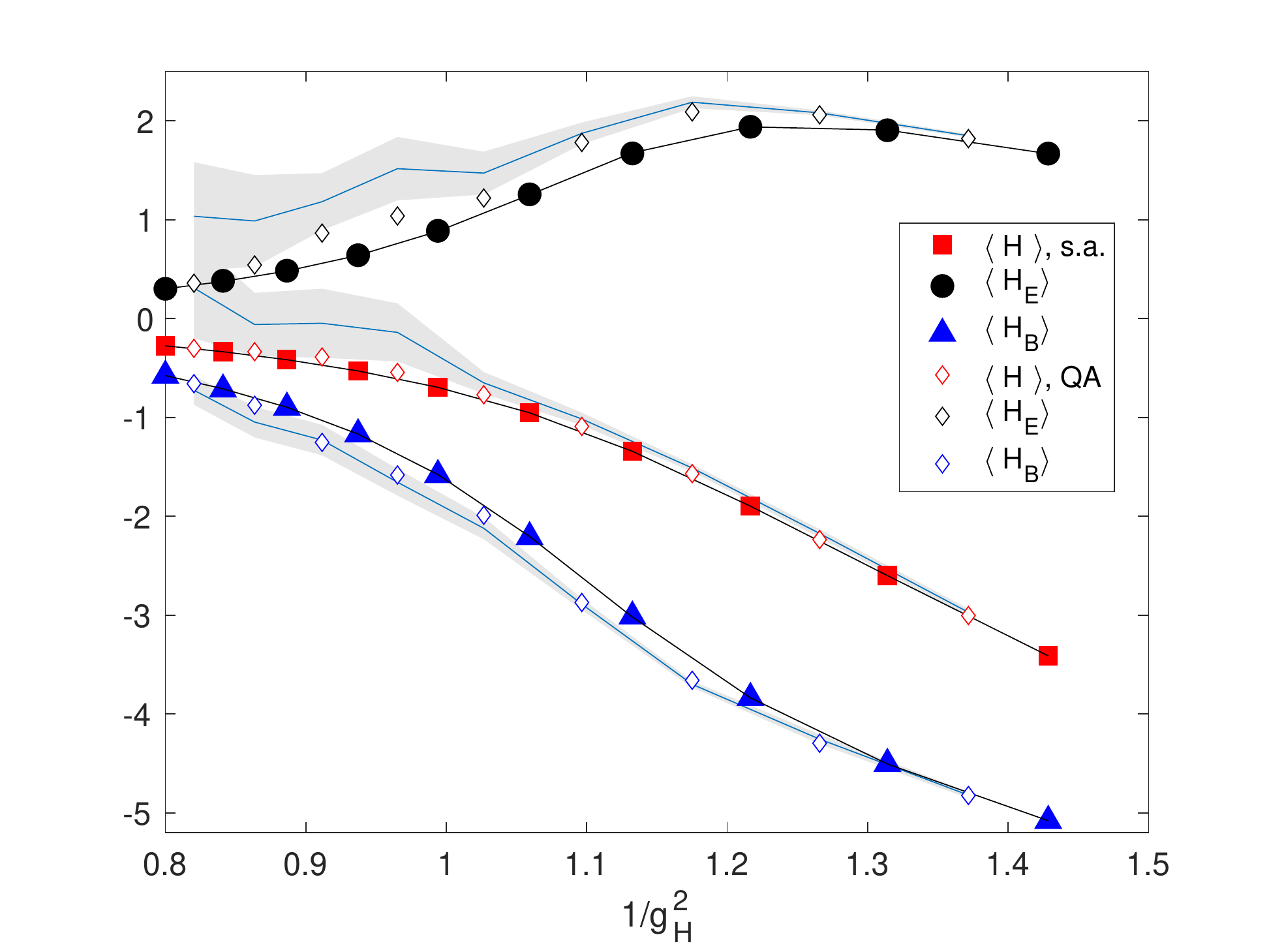}
\caption{\label{fig:D3_E_vs_gsq}(\emph{Left}): Ground-state expectation values for $D_3$ for the full Kogut-Susskind Hamiltonian (red), electric term (black), and magnetic term (blue) as a function of the inverse Hamiltonian coupling squared. The open symbols represent the minimum result from the quantum annealer while the filled symbols were obtained from classical simulated annealing. The colored band, although barely visible, represents the mean and sample standard deviation of the measurements from the quantum annealer. Simulation parameters for the latter were $K=3$ with $z_{max} = 5$ zoom steps and $n_{\mathrm{reads}} = 1000$.  (\emph{Right}): The same quantities for $G = D_4$ where the sample standard deviation from QA is much larger. This is due to the fact that more computing resources are needed to accurately determine the minimum. Simulation parameters for QA were $K=2$ with $z_{max} = 7$ zoom steps and $n_{\mathrm{reads}} = 2000$.}
\end{figure*}

where
\beq
\label{eq:qubo_groundstate}
Q_{\alpha\beta,ij} &=& 2^{i+j-2K-2z}(-1)^{\delta_{iK} + \delta_{jK}} h_{\alpha\beta} + \delta_{\alpha\beta}\delta_{ij}\tilde{Q}_{\alpha,i}\nonumber\\
\tilde{Q}_{\alpha,i} &=&  2^{i-K-z+1}(-1)^{\delta_{iK}}\sum_{\gamma}^{N_{conf}}a_{\gamma}^{(z)}h_{\gamma\alpha}\,\\
h_{\alpha\beta} &=& H_{\alpha\beta}-\eta\delta_{\alpha\beta}.\nonumber
\eeq

Here, $\eta$ represents the tunable  parameter multiplying the penalty term.
In addition to giving the variational parameters a floating-point representation, in the above definitions we have already introduced the parameter $z$, which is used in the adaptive variational search method \cite{Illa2022}. This procedure iteratively improves the estimates for the $a^{(z+1)}_\alpha$ by distributing the $K$ sampling points around the preceding solution to the Qubo problem, $a^{(z)}_\alpha$,
\beq
a_\alpha^{(z+1)} &=& a_{\alpha}^{(z)}-\frac{q_{\alpha,K}}{2^z} + \sum_{i=1}^{K-1}\frac{q_{\alpha,i}}{2^{K-i+z}}\,,
\eeq
starting at $a_\alpha^{(0)} = 0$. The estimate for the eigenstate is refined at each step, hence the name ``zooming" for this procedure. The number of zoom steps plays a significant role in the overall computational cost of our calculation, as each refinement requires calls to the quantum annealer.

The quantum computations are done on the quantum annealing hardware {\tt Advantage\_system5.1} from D-Wave~\cite{Dwave}, which is accessible via its API D-Wave Ocean~\cite{dwave_ocean}. As our system sizes are still small (c.f. Table~\ref{table:counts_configs}), the results can be compared with the exact solution using the Hamiltonian Eq.(\ref{eq:HKS_sum}) as well as with simulated annealing via the Ocean package {\tt neal}. As alternative, which we did not employ, D-Wave offers hybrid solvers s.a.~the {\tt KerberosSampler}~\cite{dwave_ocean} that attempt to break down the original QUBO matrix into smaller pieces to be subsequently solved using classical or quantum hardware. This appears to be particularly useful for system sizes that cannot be embedded on currently available annealer architectures. 

It should be noted that for computations employing both, simulated annealing and quantum annealing, results still have a $\eta$ dependence, see Eq.(\ref{eq:qubo_groundstate}). As already noticed in~\cite{Illa2022, Lewis2021}, convergence to the true ground-state is achieved for $\eta$ lying in the actual vicinity of the ground-state energy, $E_0$ (approaching from above). For practical reasons, one can determine the ``suitable" $\eta$ for a given $Q$ by solving Eq.(\ref{eq:qubo_groundstate}) iteratively in $\eta$, terminating the calculation when a certain convergence criterion, s.a. relative improvement in the solution, is fulfilled. This strategy works well for local computations with simulated annealing and could in principle be employed also for quantum annealing. Here, runtime on the quantum annealer is the major constraint.

We finally comment on our setup when accessing the annealer via the provided software package. We use the quantum annealer in its forward annealing mode with default annealing schedule and annealing time $t_f = 20~\mathrm{\mu s}$. At least one more parameter needs to be provided by the user during the quantum annealing computations. This is what is referred to as the chain strength. For our calculations, we find automatic chain strength tuning (default option) to be sufficient. Fig.\ref{fig:D3_E_vs_gsq} shows results from both, simulated annealing and quantum annealing, for the ground-state energy $\langle H\rangle$ (red) as well as the expectation value for the magnet part $\langle H_B\rangle$ (blue) and kinetic part $\langle H_E\rangle$ (black) for $G = D_3$ (left). When going to $G=D_4$ (right), an increase in computational resources is needed due to the more complicated energy landscape for the larger group, which we however only partially meet due to runtime restrictions.

\subsection{Time Evolution}
One of the main motivations for working in the Hamiltonian formulation of lattice gauge theories is the ability to access real-time dynamics. This stands in stark contrast to mainstream lattice calculations which work in Euclidean space and must perform an analytic continuation of numerical data in order to access real-time physics. In the gate-based approach, the so-called Trotter approximation is employed to the time evolution operator, $\hat{U}(T) = \exp \{ -iT \hat{H} \}$, which evolves an initial state by a finite time $T$ \cite{Lamm2019}. This approximates consists of replacing the full $\hat{U}$ with products of operators which evolve the system on a smaller time interval, $\delta t$. Corrections to this approximation typically scale with powers of $\delta t$. This approach to time-evolution of quantum states allows for an efficient simulation of the theory using universal quantum computers. 

In order to solve this problem on the quantum annealer, however, one must reformulate time evolution as an optimization problem. This can be done by the introduction of Feynman clock states \cite{Feynman:85}, a mechanism first applied to quantum chemistry calculations in order to generate parallel-in-time quantum dynamics \cite{pnas.1308069110}. We thus have to introduce an ancillary quantum system with states $\ket{t}, t=1,2,\dots, N_{t}$ where $N_t$ is the number of time-slices in the time evolution. Tensoring this orthonormal state with our as-yet-unknown state vector $\ket{\psi_t}$ at each timeslice, the problem of time evolution is equivalent to the minimization of the following functional
\beq \nn 
\mathcal{L} &=& \sum^{N_t}_{t,t'=1} \bra{t'} \bra{\psi_{t'}} \hat{\mathcal{C}} \ket{\psi_{t}} \ket {t} \\ \label{eq:time_evolution_functional} &-& \eta \left( \sum^{N_t}_{t,t'=1} \bra{t'} \braket{\psi_{t'}|\psi_{t}} \ket{t} - 1 \right),
\eeq 
where
\beq \nn 
\hat{\mathcal{C}} &\equiv& \hat{\mathcal{C}}_0 + \frac{1}{2} \bigg( \mathbb{I} \otimes \ket{t}\bra{t} + \mathbb{I} \otimes \ket{t+1}\bra{t+1} \\ \label{eq:feynman_clock_operator} &-& \hat{U}_{\delta t} \otimes \ket{t+1}\bra{t} - \hat{U}_{\delta t}^{\dagger} \otimes \ket{t}\bra{t+1} \bigg).
\eeq
\begin{figure}
\includegraphics[width=\linewidth]{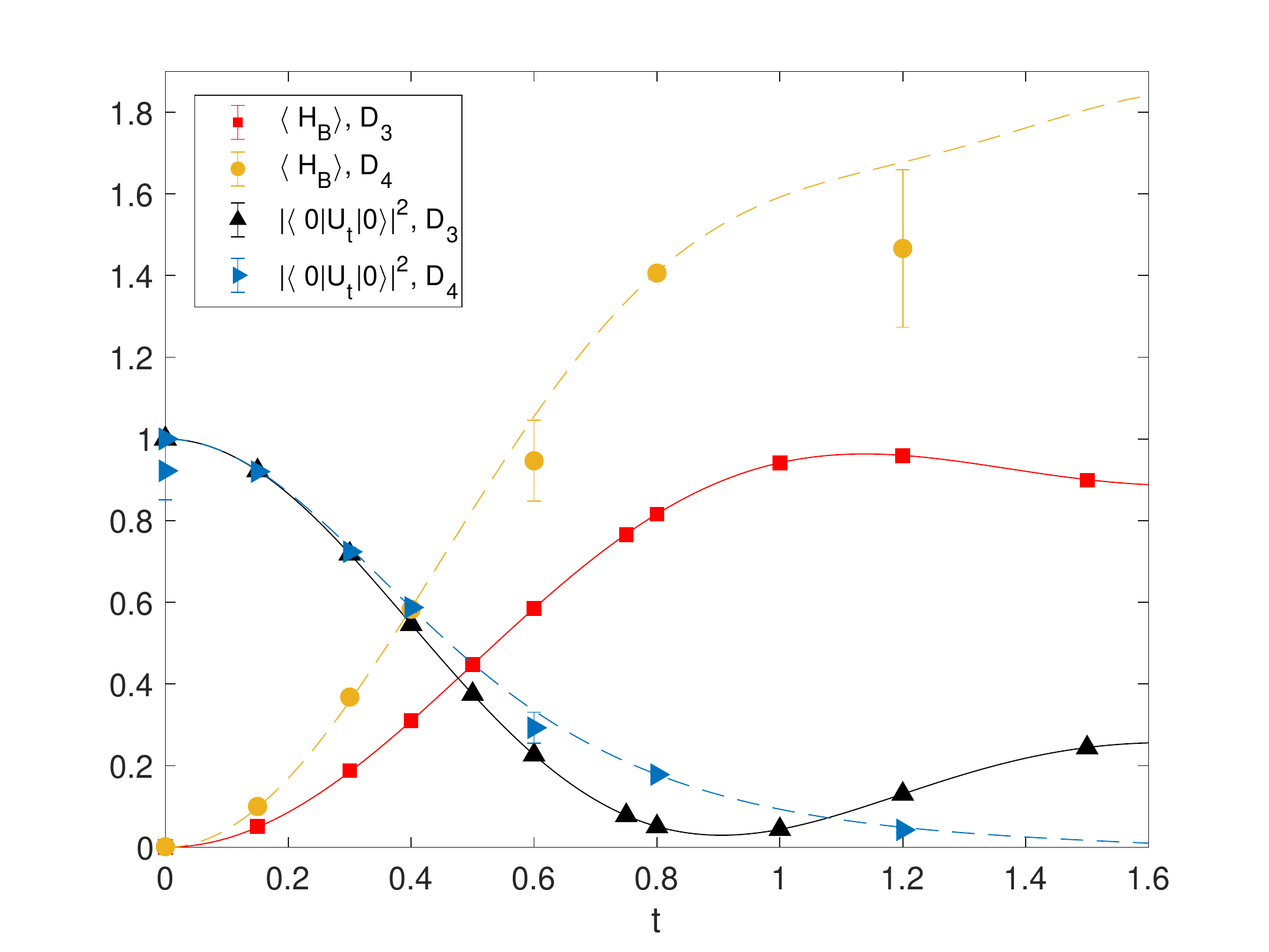}
\caption{\label{fig:DN_E_vs_gsq} Results for time evolution using simulated annealing at $g_H^2 = 0.75$ with simulation parameters given in the text. The red and yellow points represent the expectation value of the magnetic Hamiltonian in the time-evolved trivial vacuum state as a function of time for $D_3$ and $D_4$. The blue and black points represent the probability amplitude for the trivial vacuum state to persist as a function of time. The lines represent the exact result for each case.}
\end{figure}
Here $\delta t \equiv T / N_t$ is the step size in time, $\eta$ is a Lagrange multiplier analogous to our previous penalty term, and $\hat{\mathcal{C}}_0$ selects a predetermined initial state. By construction, $\hat{\mathcal{C}}$ is hermitian. One can show that the minimum of the functional Eq.(\ref{eq:time_evolution_functional}) corresponds to the exact time-evolved state at each step. Thus, as a result of a single optimization problem one obtains the the full time-evolution of a many-body quantum state over a finite time interval. 
From this functional one can now obtain the QUBO matrix. As previously discussed for the case of finding the ground state of our Hamiltonian, this is what the quantum annealer requires as input. Our discussion closely follows the derivation of \cite{SavagePreprint2022}. Using a variational state $\ket{\psi_t}\ket{t}$ at each time step $t$, the functional in Eq.(\ref{eq:time_evolution_functional}) becomes
\beq 
\mathcal{L} = \sum_{\alpha \beta} a^*_{\alpha} L_{\alpha \beta} a_{\beta},
\eeq
where the expansion parameters $a_{\alpha}$ are complex, the indices in the sum run over all $N_t N_{\mathrm{conf}}$ values, and $L_{\alpha \beta}$ are the matrix elements of the functional. The terms in the above sum can be written in terms of the real and imaginary parts of both $L_{\alpha \beta}$ and $a_{\alpha}$. Using the fixed-point representation for both the real and imaginary parts of $a_{\alpha}$, one obtains the following QUBO matrix
\begin{widetext}
\begin{equation}
    Q_{\alpha,i;\beta,j} = \begin{cases}
    2^{i+j-2K-2z} (-1)^{\delta_{iK}+\delta_{jK}}  \Re L_{\alpha\beta}+ 2\delta_{\alpha\beta} \delta_{ij} 2^{i-K-z} (-1)^{\delta_{iK}} \sum_\gamma \left( \Re a^{(z)}_\gamma  \Re L_{\gamma\beta} + \Im a^{(z)}_\gamma \Im L_{\gamma\beta}\right) & 1\leq i,j \leq K \, , \\
    - 2^{i+j'-2K-2z}(-1)^{\delta_{iK} + \delta_{j'K}} \Im L_{\alpha\beta} & 1\leq i, j' \leq K \, , \\
    2^{i'+j-2K-2z}(-1)^{\delta_{i'K} + \delta_{jK}} \Im L_{\alpha\beta} & 1\leq i',j \leq K , , \\
    2^{i'+j'-2K-2z}(-1)^{\delta_{i'K} + \delta_{j'K}}  \Re L_{\alpha\beta} + 2\delta_{\alpha\beta} \delta_{i'j'} 2^{i'-K-z} (-1)^{\delta_{i'K}} \sum_{\gamma} \left(\Im a^{(z)}_\gamma \Re L_{\gamma\beta}-  \Re a^{(z)}_\gamma \Im L_{\gamma\beta}\right) & 1\leq i',j' \leq K \, .
\end{cases}
\label{eq:qubo_matrix_time_evolution}
\end{equation}
\end{widetext}
where the Latin indices now run from $1$ to $2K$ to allow for $K$ bits in representing both the real and imaginary parts of the variational parameters and the primed Latin indices are shifted by $K$. The dimension of this QUBO matrix is $2KN_t N_{\mathrm{conf}} \times 2KN_t N_{\mathrm{conf}}$, which is significantly larger than the one used to determine the eigenstates of the Hamiltonian. This problem size is too large to be fully embedded on current quantum annealers and we revert to local simulated annealing\footnote{Simulation parameters are $K=3, N_t = 3, z_\mathrm{max}=11$ and $n_{\mathrm{runs}} = 8000$.} to solve it. 
The results of our time evolution simulations are displayed in Fig.(\ref{fig:DN_E_vs_gsq}) where we have time evolved the trivial vacuum. Shown are the expectation value of the magnetic part, $\hat{H}_B$, as well as the probability that the trivial vacuum persists, $\left|\bra{0} U(t)\ket{0}\right|^2$. One can see good agreement with the exact results.

\section{Conclusion and Outlook}
We have demonstrated how to map a simple, non-abelian lattice gauge theory onto a quantum annealer. In doing so, we have been able to compute the spectrum as well as time evolution for both $D_3$ and $D_4$ on small lattices. Furthermore, we have provided a detailed explanation of how to construct the physical states for the Kogut-Susskind Hamiltonian for an arbitrary group.
In principle, this construction could be generalized to both larger groups as well as larger lattices. The main obstacle in doing so is the scaling of the physical Hilbert space with increasing lattice and group size. It thus may be helpful to look in other directions in order to utilize the power of the quantum annealer.

One idea to get around the barrier of rapidly expanding Hilbert spaces for the case of continuous groups is to perform a truncation in the number of allowed irreps as was done in earlier studies of $SU(2)$ and $SU(3)$ in the Hamiltonian formulation~\cite{Klco2020,Klco2021,Lewis2021}. This could also be done for $D_n$, where $n>4$. A similar truncation procedure would involve both the electric and magnetic terms in the Hamiltonian and thus one could investigate the affects on the ground-state energy as well as the dynamics of the system.

One further avenue that could be pursued with the annealer is the estimation of tunneling rates and vacuum decay~\cite{PhysRevD.103.016008,Abel2020}. These non-perturbative processes are fundamental to the understanding of a wide range of phenomena in both high-energy and condensed matter physics. Another possibility is using quantum annealing in state preparation. This is an important problem faced by gate-based approaches to simulating lattice gauge theories.

\acknowledgments
The authors would like to thank Brendan Pankovich for useful discussions and Michael Spannowsky for comments on the manuscript. The work is supported by the Deutsche Forschungsgemeinschaft (DFG, German Research Foundation) through the grant CRC-TR 211 ``Strong-interaction matter under extreme conditions''~--~project number 315477589~--~TRR 211 and by the State of Hesse within the Research Cluster ELEMENTS (Project ID 500/10.006).

\bibliographystyle{ieeetr}
\bibliography{pub}

\onecolumngrid

\appendix
\section{The Gauge Group $D_n$} \label{sec:appendix_A}
In this appendix we summarize key facts and properties of the dihedral group, $D_n$, that are relevant to this work. In particular, we list the irreducible representations for general $n$, the Clebsch-Gordan coefficients for both $D_3$ and $D_4$, and outline the algorithm for constructing the Wigner $3$J-symbols.  

\subsection{Irreducible Representations}
One can define the dihedral group as the symmetry group of a regular, $n$-sided polygon. The symmetry operations are thus, rotation and reflections in the plane in which the polygon resides. The order of the group, denoted by $|D_n|$, is $2n$. We denote the group element which corresponds to rotations by $2\pi/n$ by $r$ and that which corresponds to a reflection about, say, the $y$-axis by $s$. These two elements satisfy the following relations
\beq \label{eq:dn_gen_relations}
r^n = s^2 = e, ~srs = r^{-1} = r^{n-1},
\eeq 
where $e$ denotes the identity element. Any element of $D_n$ can be uniquely represented either as $r^k, 0 \le k < n$, or as $sr^k, 0 \le k < n$. 
We now list the irreducible representations (irreps) of $D_n$ for arbitrary $n$. We recall that a representation of a group is irreducible if it contains no invariant subspaces. 

We start with the one-dimensional irreps of the dihedral group. For $n$ even, there are four one-dimensional irreps while for $n$ odd, there are two one-dimensional irreps. This is summarized by the characters listed in Table \ref{table:Dn_characters_1d}. Two-dimensional irreps of $D_n$ for arbitrary $n$ also exist and are given, in the standard basis, by
\beq 
\label{eq:two_dim_rep_standard}
\rho^h_s( r^j ) = \begin{pmatrix}
\cos \frac{2\pi hj}{n} & -\sin \frac{2\pi hj}{n} \\
\sin \frac{2\pi hj}{n} & \cos \frac{2\pi hj}{n}
\end{pmatrix}, \ \ \rho^h_s( sr^j ) = \begin{pmatrix}
-\cos \frac{2\pi hj}{n} & \sin \frac{2\pi hj}{n} \\
\sin \frac{2\pi hj}{n} & \cos \frac{2\pi hj}{n}
\end{pmatrix},
\eeq 
where we require $0 < h < n/2$, where $h$ labels the various irreps. From (\ref{eq:two_dim_rep_standard}), the characters can be easily read off 
\beq \label{eq:characters_two_dim_rep}
\chi_h(r^j) = 2 \cos \left( \frac{2\pi hj}{n} \right), ~\chi_h(sr^j) = 0.
\eeq 
These will prove useful when decomposing an arbitrary tensor representation into a direct sum of irreps. 

\begin{table}[]
\centering
\begin{tabular}{|c| c c |} 
\hline
  &$r^k$ & $sr^k$
 \\ [0.5ex] 
 \hline
 $\chi_1(g)$ & 1 & 1 \\ 
  $\chi_2(g)$ & 1 & -1 \\ 
 $\chi_3(g)$ & $(-1)^k$ & $(-1)^k$ \\ 
  $\chi_4(g)$ & $(-1)^k$ & $(-1)^{k+1}$ \\ [1ex] 
 \hline
\end{tabular}
\caption{Character table for the one-dimensional irreps of $D_n$. For $n$ even, all four irreps are valid, while for $n$ odd, only the first two rows apply. In discussing $D_3$, we will commonly refer to these one-dimensional irreps as the trivial representation, denoted by $\mathbb{1}$, and the $\sigma$-representation.}
\label{table:Dn_characters_1d}
\end{table}

\subsection{Clebsch-Gordan Coefficients}
An important task in many branches of physics is to understand how one decomposes a tensor product of two irreps into a direct sum of irreps. For the case of $D_n$ lattice gauge theory, in order to construct the physical Hilbert space by acting with gauge-invariant loop operators, one must perform exactly this task. We consider two unitary irreps of a group $G$, $\rho^p$ and $\rho^q$. In general, the tensor product of these two representations, denoted by $\rho^p \otimes \rho^q$, is reducible. This implies that it is equivalent to a direct sum of unitary irreps, which can be symbolically expressed as
\beq \label{eq:tensor_product_decomposition}
\rho^p \otimes \rho^q \approx \sum_r \oplus ~n^r_{p,q} \rho^r,
\eeq 
where $n^r_{p,q}$ is the number of times that the irrep labeled by $r$ appears in the direct sum. It is immediately clear that $n^r_{p,q}$ is symmetric in $p$ and $q$. The dihedral group $D_n$ is an example of a so-called simply reducible group, where, among other things, $n_r=0,1$ \cite{WignerCW}. The following analysis is made simpler by the requirement of only working with simply reducible groups, although it can be generalized to groups where the multiplicities can take values larger than one \cite{CornwellVanDenBroek}.

The Clebsch-Gordan (CG) coefficients can be thought of as a non-singular matrix transformation $U$, from the direct product basis to the block-diagonal basis. This is expressed by the following relation
\beq \label{eq:cg_transformation_tensor_to_irreps}
\left( \rho^p \otimes \rho^q \right) U = U \left( \sum_r \oplus ~n^r_{p,q} \rho^r \right),
\eeq 
there $U$ is of the size $d_pd_q \times d_pd_q$, where $d_r$ denotes the size of the irrep $r$. The CG coefficients are denoted by
\beq \label{eq:cg_def}
U_{i,j;r,\alpha,k} \equiv \left( \begin{array}{cc|c} p & q & r \\ i & j & k \end{array} \right),
\eeq 
where $i=1,2,\dots,d_p$, $j=1,2,\dots,d_q$, and $k=1,2,\dots,d_r$. We list here the results for $D_3$ and $D_4$, where in principle, the CG coefficients can be determined for arbitrary $n$.

\subsubsection{\texorpdfstring{$D_3$}{D3}}
The tensor products of the one-dimensional irreps are straightforward to compute as $n^{\sigma}_{\mathbb{1},\sigma} = n^{\mathbb{1}}_{\sigma,\sigma} = n^{\mathbb{1}}_{\mathbb{1},\mathbb{1}} = 1$. We list below the results for the tensor products involving the two-dimensional irrep (labeled as $\mathbb{2}$)
\beq \label{eq:CG_D3_212}
\left( \begin{array}{cc|c} 2 & \mathbb{1} & 2 \\ m &  & l \end{array} \right) = \delta_{ml}, \\ \label{eq:CG_D3_2P2}
\left( \begin{array}{cc|c} 2 & \sigma & 2 \\ m &  & l \end{array} \right) =\epsilon_{ml}, \\ \label{eq:CG_D3_221}
\left( \begin{array}{cc|c} 2 & 2 & \mathbb{1} \\ m & n & \end{array} \right) = \frac{\delta_{mn}}{\sqrt{2}}, \\ \label{eq:CG_D3_22P}
\left( \begin{array}{cc|c} 2 & 2 & \sigma \\ m & n & \end{array} \right) = \frac{\epsilon_{mn}}{\sqrt{2}}, \\ \label{eq:CG_D3_222}
\left( \begin{array}{cc|c} 2 & 2 & 2 \\ m & n & l\end{array} \right) = \frac{1}{\sqrt{2}} \left( \delta_{l2} \sigma^z_{mn}- \delta_{l1} \sigma^x_{mn} \right),
\eeq
where the indices $m,n,l=1,2$, enumerating the basis elements of the two-dimensional irrep.
\subsubsection{\texorpdfstring{$D_4$}{D4}}
With the addition of two additional one-dimensional irreps, the case of $D_4$ is a straightforward extension of the calculations for $D_3$. The tensor products of the one-dimensional irreps are summarized by
\beq \label{eq:d4_1D_tensor}
n^2_{34} = n^4_{23} = n^3_{24} = n^{q}_{1 q}  = 1, ~\tilde{q} = 1,2,3,4,
\eeq 
where we use the same labeling of the one-dimensional irreps as in Table \ref{table:Dn_characters_1d}. The CG coefficients for the $\mathbb{2}$ (two-dimensional) irrep tensored with the various one-dimensional irreps are as follows
\beq \label{eq:CG_D4_525}
\left( \begin{array}{cc|c} \mathbb{2} & 2 & \mathbb{2} \\ m &  & l\end{array} \right) = \epsilon_{ml}, \\ \label{eq:CG_D4_535}
\left( \begin{array}{cc|c} \mathbb{2} & 3 & \mathbb{2} \\ m &  & l\end{array} \right) = \sigma^z_{ml}, \\ \label{eq:CG_D4_545}
\left( \begin{array}{cc|c} \mathbb{2} & 4 & \mathbb{2} \\ m &  & l\end{array} \right) = \sigma^x_{ml}, \\ \label{eq:CG_D4_515}
\left( \begin{array}{cc|c} \mathbb{2} & 1 & \mathbb{2} \\ m &  & l\end{array} \right) = \delta_{ml}.
\eeq 
The tensor product of the $\mathbb{2}$ irrep with itself gives a direct sum of all four one-dimensional irreps. The CG coefficients associated with this decomposition are given by
\beq \label{eq:CG_D4_551}
\left( \begin{array}{cc|c} \mathbb{2} & \mathbb{2} & 1 \\ m & n & \end{array} \right) = \frac{\delta_{mn}}{\sqrt{2}}, \\ \label{eq:CG_D4_552}
\left( \begin{array}{cc|c} \mathbb{2} & \mathbb{2} & 2 \\ m & n & \end{array} \right) = \frac{\epsilon_{mn}}{\sqrt{2}}, \\ \label{eq:CG_D4_553}
\left( \begin{array}{cc|c} \mathbb{2} & \mathbb{2} & 3 \\ m & n & \end{array} \right) = \frac{\sigma^z_{mn}}{\sqrt{2}}, \\ \label{eq:CG_D4_554}
\left( \begin{array}{cc|c} \mathbb{2} & \mathbb{2} & 4 \\ m & n & \end{array} \right) = \frac{\sigma^x_{mn}}{\sqrt{2}}.
\eeq 
For $D_n, n>4$, the analysis becomes more involved as multiple two-dimensional irreps appear.
\subsection{\texorpdfstring{Wigner $3$J-symbols}{Wigner3JSymbols}}\label{subsection:3J}
In order to construct the physical Hilbert space we need to obtain the Wigner $3$J-symbols for $D_n$. Although the construction for $SU(2)$ is familiar to most, it turns out that there exists a generalization to finite groups \cite{Hu2013,PhysRevB.97.195154}.

Before providing the relevant formulae, we first provide a few key definitions from representation theory which will aid our discussion.
We start by considering a generic irrep of a finite group denoted by $(\rho, V)$, where $\rho$ is a homomorphism from the group elements to a vector space $V$. To each irrep $j$ we can associate a dual rep, denoted by $j^*$, which is equivalent to the rep $\{\rho^*(g), ~\forall g \in G\}$. The irrep $j$ is related to its dual via the so-called duality map $\Omega^j$, which is unitary and commutes with the action of the representation matrices. Furthermore, the duality map has the property that $(\Omega^{j})^T = \alpha_j \Omega^{j}$, where $\alpha_j = \pm 1,0$ is the Frobenius-Schur (FS) indicator.

We denote the 3$J$ symbol for the irreps of group $G$ labeled by the tuple $(j_1,j_2,j_3)$ by
\beq \label{eq:3J_Hu_notation}
C_{j_1j_2j_3m_1m_2m_3} \equiv \tj{j_1}{j_2}{j_3}{m_1}{m_2}{m_3},
\eeq 
where the right-hand side uses the familiar notation from $SU(2)$. The cyclic property of the $3$J symbol is given by
\beq \label{eq:3J_cyclic} 
C_{j_1j_2j_3;m_1m_2m_3} = \alpha_{j_3} C_{j_3j_1j_2;m_3m_1m_2},
\eeq 
while one can obtain the $3$J symbol for the dual of the tuple of irreps using 
\beq \label{eq:3J_dagger} 
C_{j^*_3j^*_2j^*_1;m_3m_2m_1} = \sum_{n_1n_2n_3} \overline{C_{j_1j_2j_3;n_1n_2n_3} \Omega^{j^*_3}_{n_3m_3} \Omega^{j^*_2}_{n_2m_2} \Omega^{j^*_1}_{n_1m_1}},
\eeq 
where the bar represents complex conjugation.
Finally, the normalization of the $3$J symbol is fixed by requiring 
\beq \label{eq:3J_normalization}
\sum_{\{m_i\}, \{n_i\}} \Omega^{j_3}_{m_3n_3} \Omega^{j_2}_{m_2n_2} \Omega^{j_1}_{m_1n_1}C_{j_1j_2j_3;m_1m_2m_3} C_{j^*_3j^*_2j^*_1;n_3n_2n_1}  = 1.
\eeq
This guarantees that when we couple three irreps together at a site of our ladder via the $3$J symbol, the resulting singlet state will be normalized to one. The duality map as well as the $3$J symbol can be constructed for arbitrary irreps by using the algorithm presented in \cite{Hu2013}. 
\subsection{\texorpdfstring{$D_3$}{D3}}
The results for the $3$J symbols for $D_3$ are as follows
\beq \label{eq:3J_D3_1D}
C_{\mathbb{1}\mathbb{1}\mathbb{1}} = C_{\mathbb{1}\sigma \sigma} = 1, \\ \label{eq:3J_D3_022} 
C_{\mathbb{1}\mathbb{2}\mathbb{2};m_1m_2} = \begin{pmatrix}
\frac{1}{\sqrt{2}} & 0 \\
0 & \frac{1}{\sqrt{2}}
\end{pmatrix}_{m1,m2}, \\  \label{eq:3J_D3_122} 
C_{\sigma \mathbb{2} \mathbb{2};m_1m_2} = \begin{pmatrix}
0 & \frac{i}{\sqrt{2}}  \\
\frac{-i}{\sqrt{2}} & 0
\end{pmatrix}_{m1,m2}, \\  \label{eq:3J_D3_222}
C_{\mathbb{2}\mathbb{2}\mathbb{2};m_1m_2m_3} = \begin{pmatrix}
\{ 0, \frac{1}{2} \} & \{\frac{1}{2},0 \}  \\ 
\{\frac{1}{2},0 \} & \{ 0, -\frac{1}{2} \}
\end{pmatrix}_{m1,m2,m_3}.
\eeq 
\subsection{\texorpdfstring{$D_4$}{D4}}
In a similar fashion, one can obtain the $3$J symbols for $D_4$, which read
\beq 
C_{111} = C_{122} = C_{133} = C_{144} = C_{1\mathbb{2}\mathbb{2}} = C_{34\mathbb{2}} = 1, \\ 
C_{1\mathbb{2}\mathbb{2}} = \begin{pmatrix}
\frac{1}{\sqrt{2}} & 0 \\
0 & \frac{1}{\sqrt{2}}
\end{pmatrix}_{m1,m2}, C_{2\mathbb{2}\mathbb{2}} = \begin{pmatrix}
0 & \frac{i}{\sqrt{2}}  \\
-\frac{i}{\sqrt{2}} & 0
\end{pmatrix}_{m1,m2}, \\ 
C_{3\mathbb{2}\mathbb{2}} = \begin{pmatrix}
\frac{1}{\sqrt{2}} & 0 \\
0 & -\frac{1}{\sqrt{2}}
\end{pmatrix}_{m1,m2}, C_{2\mathbb{2}\mathbb{2}} = \begin{pmatrix}
0 & \frac{1}{\sqrt{2}}  \\
\frac{1}{\sqrt{2}} & 0
\end{pmatrix}_{m1,m2}. 
\eeq 
\section{Gauss's Law}\label{section:Gauss}
For the construction of the physical Hilbert space, one must enforce Gauss's law at each lattice site. This is precisely the expression of local gauge invariance. For the more common case where compact Lie groups are employed, this condition can be written down in terms of the generators of left and right gauge transformations on the links which enter or leave a given site of the lattice. An analogous construction, which can be generalized to discrete groups such as $D_n$, uses properties of the states characterizing the various irreps. To be more concrete, at each site of our ladder we tensor together the three links which either start or end on that site while summing over the internal degrees of freedom such that it is a singlet. This has a natural expression in terms of the Wigner $3$J-symbols, and can be written as follows
\beq  \nn
\ket{00} &=& \sum_{m_{\text{I}},\tilde{m}_{\text{I}}} \ket{j_{\text{I}},m_{\text{I}};j_{\text{I}},\tilde{m}_{\text{I}}} \braket{j_{\text{I}},m_{\text{I}};j_{\text{I}},\tilde{m}_{\text{I}}|0,0}, \\ \nn
&=&  \sum_{m_{\text{I}},\tilde{m}_{\text{I}}} \sum_{m_{\text{A}}} \sum_{m_{\text{E}}} \ket{j_{\text{A}} m_{\text{A}}} \otimes \ket{j_{\text{E}} m_{\text{E}}} \otimes \ket{j_{\text{I}} m_{\text{I}}} \braket{j_{\text{A}},m_{\text{A}};j_{\text{E}},m_{\text{E}}|j_{\text{I}},\tilde{m}_{\text{I}}} \braket{j_{\text{I}},m_{\text{I}};j_{\text{I}},\tilde{m}_{\text{I}}|0,0} , \\ \nn 
&=& \sum_{m_{\text{I}},\tilde{m}_{\text{I}}} \sum_{m_{\text{A}}} \sum_{m_{\text{E}}} (-)^{f(j_{\text{A}},j_{\text{E}},j_{\text{I}};\tilde{m}_{\text{I}})} \ket{j_{\text{A}} m_{\text{A}}} \otimes \ket{j_{\text{E}} m_{\text{E}}} \otimes \ket{j_{\text{I}} m_{\text{I}}} \tj{j_{\text{A}}}{j_{\text{E}}}{j_{\text{I}}}{m_{\text{A}}}{m_{\text{E}}}{\tilde{m}_{\text{I}}} \braket{j_{\text{I}},m_{\text{I}};j_{\text{I}},\tilde{m}_{\text{I}}|0,0} , \\ \label{eq:site_singlet_ladder}
&=& \sum_{m_{\text{I}}} \sum_{m_{\text{A}}} \sum_{m_{\text{E}}} (-)^{f(j_{\text{A}},j_{\text{E}},j_{\text{I}};m_{\text{I}})} \ket{j_{\text{A}} m_{\text{A}}} \otimes \ket{j_{\text{E}} m_{\text{E}}} \otimes \ket{j_{\text{I}} m_{\text{I}}} \tj{j_{\text{A}}}{j_{\text{E}}}{j_{\text{I}}}{m_{\text{A}}}{m_{\text{E}}}{m_{\text{I}}},
\eeq 
where $f$ is a known function which has been tabulated for $D_3$ and $D_4$ and takes integer or half-integer values. In the third line we have used the fact that the CG coefficients for the projection of the tensor product of any two irreps onto the trivial representation ($\mathbb{1}$) is diagonal. The expression in (\ref{eq:site_singlet_ladder}) is valid for continuous as well as finite groups. Using the Wigner $3$J-symbols and the CG coefficients for $D_3$ and $D_4$ which were listed in appendix (\ref{sec:appendix_A}), one can produce a gauge invariant state on the ladder by taking the tensor product of the analogous expression at each site. This is represented by the expression
\beq \label{eq:generic_gauss_psi}
\ket{\psi} = \bigotimes_{s} \ket{00}_s  = \sum_{\{m_{i,1}\}}\ldots\sum_{\{m_{i,n_s}\}}\left(\bigotimes_{\ell}\ket{j_{\ell} m_{L_{\ell}} m_{R_{\ell}}}\right)\prod_s\tj{j_{\ell_{1,s}}}{j_{\ell_{2,s}}}{j_{\ell_{3,s}}}{m_{1,s}}{m_{2,s}}{m_{3,s}} (-)^{f(j_{\ell_{1,s}},j_{\ell_{2,s}},j_{\ell_{3,s}};m_{3,s})},
\eeq
where the $\{m_{i,s}\}$ are the local multiplicities at site $s$ of the link $\ell$ in representation $j_{\ell}$, and in the product over links, each link has multiplicities $m_{L_{\ell}}$ and $m_{R_{\ell}}$ associated with the ``left" and ``right" ends of the link. One can thus label the states of the Hilbert space by a set of $N_{\text{links}} = 3N$ integers characterizing the irreps, where $N$ is the number of plaquettes, with the condition that at each site there is a valid tuple with nonzero $3$J symbol. From here, one has everything one needs in order to enumerate the states in the physical Hilbert space.
\section{Plaquette Contribution} \label{sec:plaq_contrib}
Here we provide the details for the matrix element of a plaquette in the physical Hilbert space. This derivation is similar to that of \cite{Lewis2021} for the case of $SU(2)$.
The application of the plaquette in (\ref{eq:arb_plaquette_ladder}) on the arbitrary physical state listed in (\ref{eq:generic_gauss_psi}) is given by
\beq 
\hat{P}_{\vec{x}}\ket{\psi} &=& \sum_{n_1,\ldots,n_4}\sum_{\{m_{L_i}\}}\sum_{\{m_{R_i}\}} \hat{U}_{\tilde{l}_1;n_1n_2}\ket{j_{\tilde{l}_1}m_{L_{\tilde{l}_1}}m_{R_{\tilde{l}_1}}}\\ \nn &\otimes& \hat{U}_{\tilde{l}_2;n_2n_3}\ket{j_{\tilde{l}_2}m_{L_{\tilde{l}_2}}m_{R_{\tilde{l}_2}}} \otimes  \hat{U}_{\tilde{l}_3;n_4n_3}\ket{j_{\tilde{l}_3}m_{L_{\tilde{l}_3}}m_{R_{\tilde{l}_3}}}\\ \nn &\otimes& \hat{U}_{\tilde{l}_4;n_1n_4}\ket{j_{\tilde{l}_4}m_{L_{\tilde{l}_4}}m_{R_{\tilde{l}_4}}} \bigotimes_{i \not\in \mathcal{L}_{\vec{x}}} \ket{j_i m_{L_i}m_{R_i}} \\ \label{eq:arb_plaquette_on_bra_1}
&\prod_s & \tj{j_{\ell_{1,s}}}{j_{\ell_{2,s}}}{j_{\ell_{3,s}}}{m_{1,s}}{m_{2,s}}{m_{3,s}} (-)^{f(j_{\ell_{1,s}},j_{\ell_{2,s}},j_{\ell_{3,s}};m_{3,s})}\,,
\eeq 
where the sums over $m_{L_i}$ and $m_{R_i}$ represent $3N$ individual sums whose ranges depend on the configuration of the state $\ket{\psi}$. Using the Clebsch-Gordan series, we can obtain the result of each link operator acting on an arbitrary link state. Applying this to each term in (\ref{eq:arb_plaquette_on_bra_1}) yields
\beq \nn 
\hat{P}_{\vec{x}}\ket{\psi} = \sum_{n_1,\ldots,n_4}\sum_{\{m_{L_i}\}}\sum_{\{m_{R_i}\}}\prod_s\tj{j_{\ell_{1,s}}}{j_{\ell_{2,s}}}{j_{\ell_{3,s}}}{m_{1,s}}{m_{2,s}}{m_{3,s}} (-)^{f(j_{\ell_{1,s}},j_{\ell_{2,s}},j_{\ell_{3,s}};m_{3,s})} \bigotimes_{i \not\in \mathcal{L}_{\vec{x}}} \ket{j_i m_{L_i}m_{R_i}}\otimes\\ \nn 
\sum_{J_{\tilde{l}_1}}\sum_{M_{L_{\tilde{l}_1}}}\sum_{M_{R_{\tilde{l}_1}}} \sqrt{\frac{\mathrm{dim}(j_{\tilde{l}_1})}{\mathrm{dim}(J_{\tilde{l}_1})}}\braket{2 n_1 j_{\tilde{l}_1} m_{L_{\tilde{l}_1}} | J_{\tilde{l}_1} M_{L_{\tilde{l}_1}}}\braket{J_{\tilde{l}_1} M_{R_{\tilde{l}_1}} | 2 n_2 j_{\tilde{l}_1} m_{R_{\tilde{l}_1}} }\ket{J_{\tilde{l}_1} M_{L_{\tilde{l}_1}} M_{R_{\tilde{l}_1}}}\otimes\\ \nn 
\sum_{J_{\tilde{l}_2}}\sum_{M_{L_{\tilde{l}_2}}}\sum_{M_{R_{\tilde{l}_2}}} \sqrt{\frac{\mathrm{dim}(j_{\tilde{l}_2})}{\mathrm{dim}(J_{\tilde{l}_2})}}\braket{2 n_2 j_{\tilde{l}_2} m_{L_{\tilde{l}_2}} | J_{\tilde{l}_2} M_{L_{\tilde{l}_2}}}\braket{J_{\tilde{l}_2} M_{R_{\tilde{l}_2}} | 2 n_3 j_{\tilde{l}_2} m_{R_{\tilde{l}_2}} }\ket{J_{\tilde{l}_2} M_{L_{\tilde{l}_2}} M_{R_{\tilde{l}_2}}}\otimes\\ \nn 
\sum_{J_{\tilde{l}_3}}\sum_{M_{L_{\tilde{l}_3}}}\sum_{M_{R_{\tilde{l}_3}}} \sqrt{\frac{\mathrm{dim}(j_{\tilde{l}_3})}{\mathrm{dim}(J_{\tilde{l}_3})}}\braket{2 n_4 j_{\tilde{l}_3} m_{L_{\tilde{l}_3}} | J_{\tilde{l}_3} M_{L_{\tilde{l}_3}}}\braket{J_{\tilde{l}_3} M_{R_{\tilde{l}_3}} | 2 n_3 j_{\tilde{l}_3} m_{R_{\tilde{l}_3}} }\ket{J_{\tilde{l}_3} M_{L_{\tilde{l}_3}} M_{R_{\tilde{l}_3}}}\otimes\\ \label{eq:arb_plaquette_on_bra_2}
\sum_{J_{\tilde{l}_4}}\sum_{M_{L_{\tilde{l}_4}}}\sum_{M_{R_{\tilde{l}_4}}} \sqrt{\frac{\mathrm{dim}(j_{\tilde{l}_4})}{\mathrm{dim}(J_{\tilde{l}_4})}}\braket{2 n_1 j_{\tilde{l}_4} m_{L_{\tilde{l}_4}} | J_{\tilde{l}_4} M_{L_{\tilde{l}_4}}}\braket{J_{\tilde{l}_4} M_{R_{\tilde{l}_4}} | 2 n_4 j_{\tilde{l}_4} m_{R_{\tilde{l}_4}} }\ket{J_{\tilde{l}_4} M_{L_{\tilde{l}_4}} M_{R_{\tilde{l}_4}}}\,.
\eeq 
We now take the inner product of (\ref{eq:arb_plaquette_on_bra_2}) with a second generic state $\ket{\psi'}$ which satisfies Gauss' law. We note that states in the representation basis are orthonormal
\beq \label{eq:representation_basis_orthonormality}
\braket{j'm'n'|jmn} = \delta_{jj'} \delta_{mm'} \delta_{nn'},
\eeq
and thus we obtain the result in Eq.(\ref{eq:arb_plaquette_matrix_element}). 

\end{document}